\documentclass[aps,prl,twocolumn,superscriptaddress]{revtex4}

\usepackage{dcolumn}
\usepackage{bm}

\usepackage{epsfig}
\usepackage{graphicx}
\usepackage{color}
\usepackage{amsmath,amssymb}

\newcommand{\beq}[1]{
\begin{equation}
\label{e#1} }

\newcommand{\eeq}{
\end{equation}
}

\begin{document}

\title{Spin gating electrical current}
\author{C.~Ciccarelli}
\affiliation{Cavendish Laboratory, University of Cambridge, CB3 0HE, United Kingdom}

\author{L.~P.~Z\^{a}rbo}
\affiliation{Institute of Physics ASCR, v.v.i., Cukrovarnick\'a 10, 162 53 Praha
6, Czech Republic}

\author{A.~C.~Irvine}
\affiliation{Cavendish Laboratory, University of Cambridge, CB3 0HE, United Kingdom}

\author{R.~P.~Campion}
\affiliation{School of Physics and
Astronomy, University of Nottingham, Nottingham NG7 2RD, United Kingdom}

\author{B.~L.~Gallagher}
\affiliation{School of Physics and
Astronomy, University of Nottingham, Nottingham NG7 2RD, United Kingdom}

\author{J.~Wunderlich}
\affiliation{Hitachi Cambridge Laboratory, Cambridge CB3 0HE, United Kingdom}
\affiliation{Institute of Physics ASCR, v.v.i., Cukrovarnick\'a 10, 162 53
Praha 6, Czech Republic}

\author{T.~Jungwirth}
\affiliation{Institute of Physics ASCR, v.v.i., Cukrovarnick\'a 10, 162 53
Praha 6, Czech Republic} \affiliation{School of Physics and
Astronomy, University of Nottingham, Nottingham NG7 2RD, United Kingdom}

\author{A.~J.~Ferguson}
\email{ajf1006@cam.ac.uk}
\affiliation{Cavendish Laboratory, University of Cambridge, CB3 0HE, United Kingdom}

\date{\today}

%

\begin{abstract}
We use an aluminium single electron transistor with a magnetic gate to directly quantify the chemical potential anisotropy of GaMnAs materials. Uniaxial and cubic contributions to the chemical potential anisotropy are determined from field rotation experiments. In performing magnetic field sweeps we observe additional isotropic magnetic field dependence of the chemical potential which shows a non-monotonic behavior. The observed effects are explained by calculations based on the $\mathbf{k}\cdot\mathbf{p}$ kinetic exchange model of ferromagnetism in GaMnAs. Our device inverts the conventional approach for constructing spin transistors: instead of spin-transport controlled by ordinary gates we spin-gate ordinary charge transport.
\end{abstract}
\maketitle

Single electron transistors (SETs) \cite{Likharev:1999_a} represent the ultimate in transistor miniaturization and charge sensitivity. Their transport channel comprises a small island in tunnel contact with the leads and, due to the Coulomb blockade effect, transport of electrical current through the SET takes place one electron at a time.  Spin phenomena and functionalities have been incorporated in the transport channel of SETs, both in the leads and/or in the island. Observed effects include spin accumulation on the island and large tunnelling magnetoresistances \cite{Dempsey:2011_a}. The motivation for our present work comes from studies of the large magnetoresistances due to magneto-Coulomb oscillations \cite{Ono:1997_a,Deshmukh:2002_a,vanderMolen:2006_a} or Coulomb blockade anisotropic magnetoresistance (AMR) \cite{Wunderlich:2006_a,Schlapps:2009_a,Bernand-Mantel:2009_a}, associated with chemical potential effects in single electron transistors with ferromagnetic leads or islands. These studies showed that transport through the channel can be controlled by shifts of the chemical potentials at individual components of the SET channel, induced by the Zeeman coupling to the external magnetic field \cite{Ono:1997_a,Deshmukh:2002_a,vanderMolen:2006_a} or by magnetization rotation and relativistic spin-orbit coupling \cite{Wunderlich:2006_a,Schlapps:2009_a,Bernand-Mantel:2009_a}.  Particularly intriguing is the latter phenomenon which can yield low-field hysteretic magnetoresistances of huge magnitudes and which is a direct relative of the AMR in conventional ohmic \cite{Thomson:1857_a,McGuire:1975_a} or tunneling \cite{Gould:2004_a} devices (see supplementary information). The chemical potential anisotropy has also been observed in GaMnAs tunnel devices \cite{Tran:2009_a}.

In this Letter we measure the magnetisation and field dependence of a ferromagnetic SET with no ferromagnetic material in the transport channel, instead placing it in the electrostatic gate. This spin-gating configuration, in which the transport channel is detached from the magnetic gate, allows the elimination of unwanted side effects such as chemical potential shifts due to spin accumulation or tunnelling AMR in the SET island. It allows us to directly measure the chemical potential change of the gate without the application of scaling factors. Moreover, the measured chemical potentials depend on the material rather than the specific properties of a nanodevice.

\begin{figure*}[]
\centering
\includegraphics[angle=0,width=0.9\textwidth]{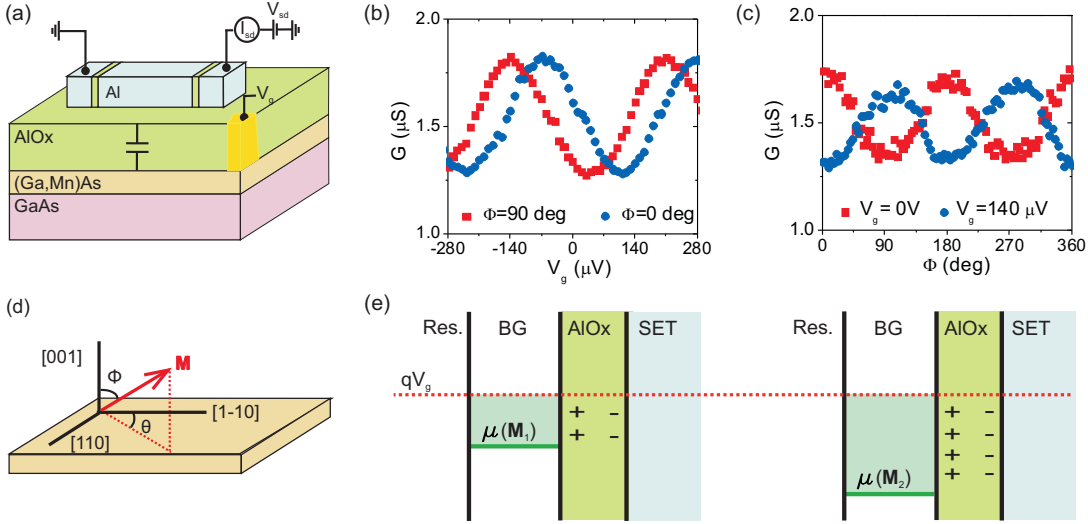}
\caption{(a), Schematic showing our SET channel separated by AlO$_x$ dielectric from the ferromagnetic GaMnAs back-gate. The SET comprises Al leads and island, and AlO$_x$ tunnel barriers (micrograph in Supplementary information). (b), Coulomb oscillations for the SET on Ga$_{0.97}$Mn$_{0.03}$As for two different polar angles $\Phi$ of the magnetisation. (c), Magneto-Coulomb oscillations shown by the same SET by varying the angle of magnetisation for two different gate voltages.  Measurements were performed using a low frequency lock-in technique with excitation voltage 20~$\mu V$ and zero dc bias. (d) Magnetisation vector with respect to GaMnAs crystal axes. (e), Schematic explaining the spin gating phenomenon: reorientation of the magnetisation from \textbf{M$_1$} to \textbf{M$_2$} causes a change in the chemical potential of the GaMnAs back-gate (BG). This causes charge to flow onto the back-gate  from the reservoir (Res.). The net effect is to alter the charge on the back-gate and therefore the SET conductance. We show a decrease in hole chemical potential $\mu$ between \textbf{M$_1$} and \textbf{M$_2$}. The externally applied electrochemical potential on the gate $\mu_{ec}=qV_g$ is held constant.
}\label{figone}
\end{figure*}

Our SET has a micron-scale aluminium island separated by aluminium oxide tunnel junctions from source and drain leads (Fig.~\ref{figone}(a)). The aluminium is lightly doped with manganese to suppress superconductivity in our low-temperature (300~mK) measurements. The SETs are fabricated on top of epitaxially grown GaMnAs layers, which are electrically insulated from the SET by an alumina dielectric, and act as a spin-back-gate to the SET. We choose GaMnAs for the gate electrode due to its well established relativistic magnetic anisotropy characteristics \cite{Jungwirth:2006_a,Dietl:2008_b}.The low total capacitance ($C_\Sigma\sim0.6$~fF) of the island to its leads and back-gate yields a single-electron charging energy ($E_c=e^2/2C_\Sigma$) of $\sim100$~$\mu$eV.

By sweeping the externally applied potential to the SET gate ($V_g$) we obtain the conductance oscillations that characterize Coulomb blockade, as shown in Fig.~\ref{figone}(b). In samples with magnetic gates we see a shift of these oscillations by an applied saturating magnetic field which rotates the magnetization in the GaMnAs gate. In Fig.~\ref{figone}(b) we show measurements for the in-plane ($\Phi=90^\circ$) and for the perpendicular-to-plane ($\Phi=0^\circ$) directions of magnetization. Alternatively, we plot in Fig.~\ref{figone}(c) the channel conductance as a function of the magnetization angle $\Phi$ for a fixed external potential $V_g$ applied to the gate. The oscillations in $\Phi$ seen in Fig.~\ref{figone}(c) are of comparable amplitude as the oscillations in $V_g$ in Fig.~\ref{figone}(b).

Due to the spin-orbit coupling, the band structure of GaMnAs is perturbed when its magnetisation \textbf{M} is rotated. One consequence is a shift of the chemical potential, which in itself doesn't yield a response of the SET. However, the back-gate is attached to a charge reservoir so any change in the internal chemical potential of the gate causes an inward, or outward, flow of charge in the gate, as illustrated in Fig.~\ref{figone}(e). This change in back-gate charge offsets the Coulomb oscillations (Fig.~\ref{figone}(b)) and changes the conductance of the transistor channel for a fixed external potential applied to the gate (Fig.~\ref{figone}(c)). Any magneto-Coulomb effect from a \textbf{M}-dependent change in the depletion region at the surface \cite{Pappert:2006_a} is ruled out: this would lead to the shift in Coulomb oscillations being dependent on the magnitude of the gate voltage.

\begin{figure}[]
\includegraphics[angle=0,width=0.5\textwidth]{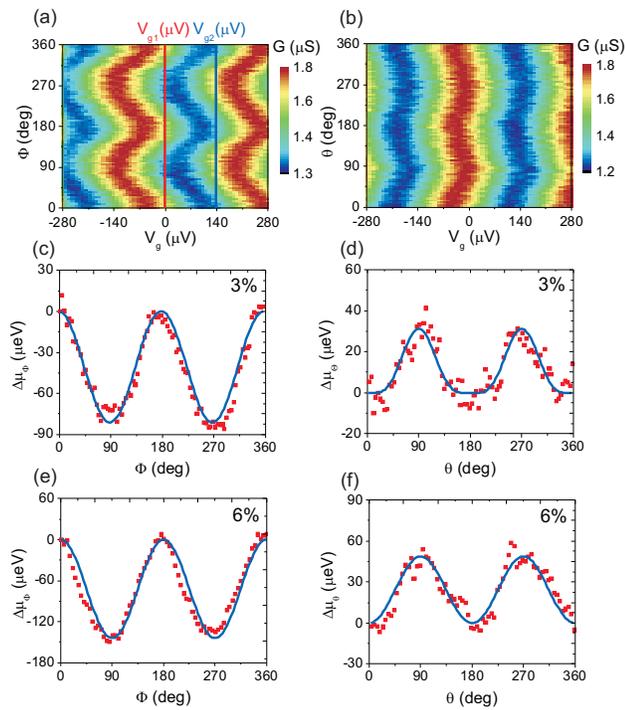}
\caption{(a), SET conductance measurements for azimuthal angle $\theta=0$ as a function of the out-of-plane polar angles $\Phi$ of the Ga$_{0.97}$Mn$_{0.03}$As magnetisation and of the gate voltage. The magnetisation is rotated by a 1 T magnetic field.  The two lines represent the gate voltage offsets for the data shown in Fig.~1c. (b), Measurements with in-plane rotation of magnetisation for the same sample. The magnetisation is rotated by a 0.8 T magnetic field. (c), Ga$_{0.97}$Mn$_{0.03}$As hole chemical potential shift  ($\Delta\mu_\Phi=\mu(\Phi)-\mu(\Phi=0)$) for the out-of-plane  magnetisation rotation, inferred from measurements in (a). (d), Ga$_{0.97}$Mn$_{0.03}$As hole chemical potential shift ($\Delta\mu_\theta=\mu(\theta)-\mu(\theta=0)$)  for the in-plane magnetisation rotation, inferred from measurements in (b). (e-f), Hole chemical potential shifts for the Ga$_{0.94}$Mn$_{0.06}$As gate electrode. The data (red squares) are fitted (blue lines) to extract the uniaxial and cubic contributions to the $\Delta\mu$ anisotropy.
}\label{figtwo}
\end{figure}

We now display the full experimental data sets for in-plane and out-of-plane magnetization rotations in a saturating magnetic field in the case of a ferromagnetic p-type Ga$_{0.97}$Mn$_{0.03}$As gate (Fig.~\ref{figtwo}(a,b)). We fit a sinusoid to the Coulomb oscillations and extract the resulting gate-voltage offset ($\Delta V_g$) as a function of the magnetization angle. A positive value of $\Delta V_g$ means that, for a fixed gate voltage, holes leave the gate, which can be attributed to an increased hole chemical potential, $\Delta\mu$=$q\Delta V_g$.

The anisotropy of $\Delta\mu$ for out-of-plane rotation of \textbf{M} has a uniaxial symmetry ($\Delta\mu_\Phi^u=$81~$\mu$eV) (Fig.~\ref{figtwo}(c)), whereas the in-plane rotation anisotropy is predominantly uniaxial ($\Delta\mu_\theta^u=30$~$\mu$eV) with a small cubic component ($\Delta\mu_\theta^c=$6$~\mu$eV) (Fig.~\ref{figtwo}(d)). The $\Delta\mu$ anisotropy is greater for the out-of-plane rotation, and causes a shift in the Coulomb oscillations by a quarter of a period. We also measured SETs with a Ga$_{0.94}$Mn$_{0.06}$As back-gate (Figs.~\ref{figtwo}(e,f)). The magnitude of both the out-of-plane rotation ($\Delta\mu_\Phi^u=144$~$\mu$eV) and the in-plane rotation ($\Delta\mu_\theta^u=46$~$\mu$eV) uniaxial anisotropies increase in the sample with higher Mn concentration while the in-plane cubic anisotropy component is negligibly small in the Ga$_{0.94}$Mn$_{0.06}$As sample. The chemical potential anisotropy constants are consistent between samples fabricated from the same MBE grown wafers. This is because the spin gating phenomenon measures the properties of the material surface (over a $\sim\mu m^2$ area in our sample) rather than a specific nanodevice.

\begin{figure}[]
\includegraphics[angle=0,width=0.5\textwidth]{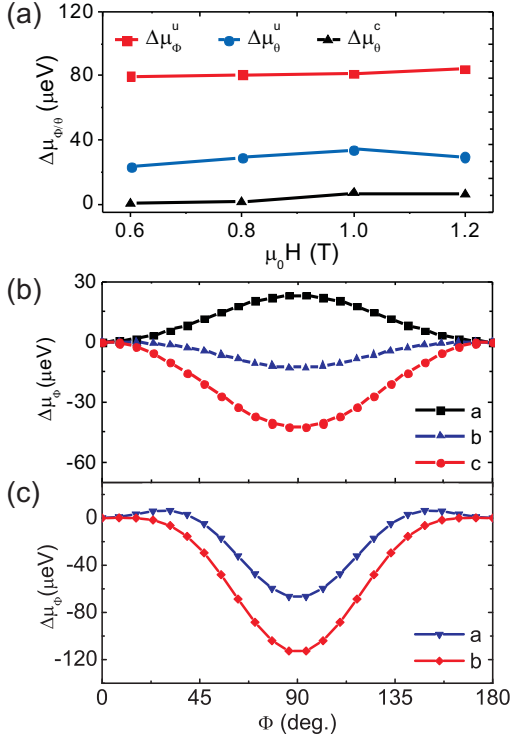}
\caption{(a), Experimental values for cubic and uniaxial contributions to the $\Delta\mu$ anisotropy, measured for different saturation fields, for the Ga$_{0.97}$Mn$_{0.03}$As gate material. (b), Theoretical variations of the chemical potential with respect to the magnetization angle $\Phi$ for fixed Mn moment density $N_{\rm Mn}=4.4\times 10^{20}{\rm cm}^{-3}$ (corresponding approximately to the measured sample with 3\% nominal doping) and compressive growth strain $\epsilon_0=-0.3$\%. The hole density $p=N_{\rm Mn}$, $p=0.4N_{\rm Mn}$ and $p=0.2N_{\rm Mn}$ for curves a, b, and c, respectively. (c), Theoretical chemical potential anisotropies for $N_{\rm Mn}=8.8\times 10^{20}{\rm cm}^{-3}$ (corresponding approximately to the measured sample with 6\% nominal doping) and charge density $p=0.2N_{\rm Mn}$. The curve a is for the strain $\epsilon_0=-0.3$\% while curve b is for $\epsilon_0=-0.5$\%.}
\label{figthree}
\end{figure}

We repeated field rotation measurements for different values of the saturating field, observing only a weak dependence of the $\Delta\mu$ anisotropy components on the field magnitude (Fig.~\ref{figthree}(a)). This confirms that the origin of the observed spin-gating is the anisotropy of the chemical potential in the ferromagnetic gate with respect to the magnetization orientation. (Recall that the effective g-factor in the GaAs valence band has a negligible anisotropy \cite{Winkler:2000_a}).

The chemical potential anisotropy origin of the observed spin-gating effect is confirmed by microscopic calculations based on the $\mathbf{k}\cdot\mathbf{p}$ kinetic-exchange model of GaMnAs \cite{Dietl:2001_b,Abolfath:2001_a} (see Figs.~\ref{figthree}(b,c) and Supplementary information). We have calculated chemical potential anisotropies for the out-of-plane magnetization rotation assuming local Mn moment concentrations and lattice-matching growth strains corresponding to the measured GaMnAs samples. We consider one hole per Mn$_{\rm Ga}$ local moment (Mn$_{\rm Ga}$ is a single acceptor), as well as cases with partial hole depletion. The latter calculations were performed since the spin-gating experiments are sensitive to the partially depleted surface layer which has a width given by the hole screening length (a few nm).

In agreement with experiment, we find theoretical chemical potential anisotropies of the order of 10-100~$\mu$eV. The calculated enhancement of the chemical potential anisotropy with increasing local moment concentration and lattice-matching growth strain is also consistent with the enhanced spin-gating effect observed experimentally in the Ga$_{0.94}$Mn$_{0.06}$As sample as compared to the Ga$_{0.97}$Mn$_{0.03}$As sample. We also note that the increase in the higher Mn-doped sample of the measured and calculated chemical potential anisotropy for the out-of-plane rotation is consistent with the doping trends in the corresponding magnetocrystalline anisotropy coefficient of GaMnAs \cite{Zemen:2009_a}. Similarly, the measured enhancement of the in-plane uniaxial component of the chemical potential anisotropy and the suppression of the cubic component in the higher Mn-doped sample is consistent with the corresponding magnetocrystalline anisotropy trends \cite{Zemen:2009_a}. The microstructural origin of the in-plane uniaxial anisotropy component is not established and to capture this component one can, e.g., introduce an effective shear strain in the $\mathbf{k}\cdot\mathbf{p}$ Hamiltonian which breaks the [110]/[1-10] crystal symmetry.\cite{Sawicki:2004_a,Zemen:2009_a} The values of the shear strain required to reproduce the in-plane uniaxial magnetocrystalline anisotropy yield a corresponding chemical potential anisotropy of the order of $\sim 10-100$~$\mu$eV, again consistent with our measured data. Similar to the established phenomenology of the spin-orbit coupling induced magnetocrystalline anisotropies, or ohmic and tunneling AMRs, the sign and magnitude of the chemical potential anisotropy is not strictly determined by the sign of the strain and by other symmetries of the crystal, and may vary when changing the local moment or hole concentration in GaMnAs. We note that the agreement in the sign of the theoretical and experimental chemical potential anisotropy is obtained assuming partially depleted holes, consistent with the expected sensitivity in our device to the GaMnAs surface layer.

\begin{figure}[]
\includegraphics[angle=0,width=0.5\textwidth]{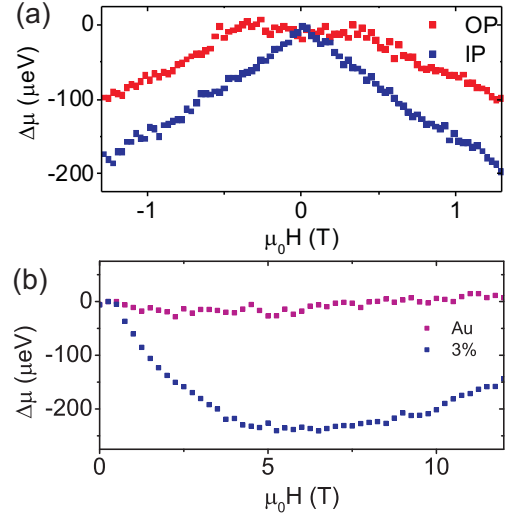}
\caption{(a), The chemical potential shift determined for out-of-plane and in plane easy axis sweeps of the magnetic field for the Ga$_{0.97}$Mn$_{0.03}$As gate material. We refer the chemical potential to its zero field value. (b), The chemical potential shifts for a high field out of plane sweep for Ga$_{0.97}$Mn$_{0.03}$As and also for a control sample with an Au gate.}
\label{figfour}
\end{figure}

In ferromagnets, the chemical potential can also shift when the magnetization angle is fixed while the magnitude of the external magnetic field is varied. Performing a sweep of the magnetic field in the out-of-plane direction, we observe a nearly flat region followed by a decreasing chemical potential (Fig.~\ref{figfour}(a)). The flat, low-field region occurs as \textbf{M} rotates from the in-plane easy axes to the out-of-plane direction. During the rotation of \textbf{M}, the \textbf{M}-orientation dependent and field-dependent contributions to $\Delta\mu$ nearly cancel. If we sweep the field along an in-plane easy axis, just the field-dependent $\Delta\mu$ is observed (Fig.~\ref{figfour}(a)), consistent with the absence of \textbf{M} rotation.

At higher fields the decreasing chemical potential saturates and eventually increases (Fig.~\ref{figfour}(b)). The decrease in chemical potential is due to an indirect mechanism in which the field acts on the hole bands via the kinetic-exchange coupling (see Supplementary information). We assume that, apart from the ferromagnetic Mn moments being saturated by low magnetic fields corresponding to the magnetic anisotropy fields, a fraction of the Mn local moment magnetization is not fully saturated at the low-fields. Sweeping the field gradually polarizes these unsaturated moments which in our theoretical modeling corresponds to an increase of the effective concentration of the ferromagnetic local moments. The hole chemical potential decreases because the enhanced strength of the kinetic-exchange field due to the larger density of saturated local moments causes an enhancement of the spin-splitting of the hole bands. The field-dependent $\Delta\mu$ saturates at a value of $\sim 200$~$\mu$eV. From the calculations we infer that such a shift corresponds to an effective $\sim1$\% increase of the density of ferromagnetic local moments. The field-induced additional polarization of a small fraction of the local moment density can be readily expected for the studied samples but would be unobservable by standard magnetisation measurements.

The mechanism for the increase in chemical potential at higher fields is the direct Zeeman coupling of the magnetic field to the hole bands (see Supplementary information). Similar to the above indirect mechanism, the sign of the chemical potential shift due to the Zeeman coupling is given by the sense in which the spin-splitting of the hole bands changes in the applied field. However, unlike the indirect mechanism, the Zeeman coupling reduces the spin-splitting of the hole bands which yields an increase of the chemical potential. The chemical potential increases with magnetic field because the magnetization of the local Mn moments and of the holes are antiparallel, a consequence of the p-d hybridization origin of the kinetic-exchange interaction \cite{Jungwirth:2006_a}. To assess quantitatively the chemical potential shift due to the Zeeman coupling we performed the $\mathbf{k}\cdot\mathbf{p}$ kinetic-exchange calculations (including spin-orbit coupling) in the presence of the magnetic field directly acting on the hole bands. As expected, the calculated shift of the chemical potential is larger for larger hole depletion and in the more depleted case we obtain an increase of the chemical potential of the order of $\sim 10$~$\mu$eV/T which is consistent with the experimental high-field trend seen in Fig.~\ref{figfour}(b). The experimental data in Fig.~\ref{figfour}(b) are shown for an out-of-plane field-sweep in Ga$_{0.97}$Mn$_{0.03}$As, however, the same behaviour is observed for the in-plane sweep, consistent with the negligible anisotropy of the Zeeman effect in the 3D valence band of GaMnAs. A control SET with a Au gate (Fig.~\ref{figfour}(b)) shows no systematic effect due to the magnetic field, and confirms that there is no significant magneto-conductance due to the aluminium-manganese alloy.

The spin-gating technique was employed to accurately measure the anisotropic, and isotropic, chemical potential phenomena in GaMnAs. However, this technique can be applied to catalogue these effects in other magnetic materials by the simple step of exchanging the gate electrode. Finally, we point out that our work comprises a novel spin transistor where the charge state of the device channel is sensitive to the spin state of its magnetic gate (see Supplementary information).

\begin{acknowledgments}
We acknowledge support from EU Grants FP7-214499 NAMASTE, FP7-215368 SemiSpinNet, ERC Advanced Grant, from Czech Republic Grants AV0Z10100521, KAN400100652, KJB100100802 and Preamium Academiae, A.J.F. acknowledges the support of a Hitachi research fellowship.
\end{acknowledgments}


\begin{thebibliography}{23}
\expandafter\ifx\csname natexlab\endcsname\relax\def\natexlab#1{#1}\fi
\expandafter\ifx\csname bibnamefont\endcsname\relax
  \def\bibnamefont#1{#1}\fi
\expandafter\ifx\csname bibfnamefont\endcsname\relax
  \def\bibfnamefont#1{#1}\fi
\expandafter\ifx\csname citenamefont\endcsname\relax
  \def\citenamefont#1{#1}\fi
\expandafter\ifx\csname url\endcsname\relax
  \def\url#1{\texttt{#1}}\fi
\expandafter\ifx\csname urlprefix\endcsname\relax\def\urlprefix{URL }\fi
\providecommand{\bibinfo}[2]{#2}
\providecommand{\eprint}[2][]{\url{#2}}

\bibitem[{\citenamefont{Likharev}(1999)}]{Likharev:1999_a}
\bibinfo{author}{\bibfnamefont{K.~K.} \bibnamefont{Likharev}},
  \bibinfo{journal}{Proceedings of the IEEE} \textbf{\bibinfo{volume}{87}},
  \bibinfo{pages}{606} (\bibinfo{year}{1999}).

\bibitem[{\citenamefont{Dempsey and Marrows}(2011)}]{Dempsey:2011_a}
\bibinfo{author}{\bibfnamefont{K.~J.} \bibnamefont{Dempsey}} \bibnamefont{and}
  \bibinfo{author}{\bibfnamefont{D.~C. C.~H.} \bibnamefont{Marrows}},
  \bibinfo{journal}{Phil. Trans. R. Soc.} \textbf{\bibinfo{volume}{A 369}},
  \bibinfo{pages}{3150} (\bibinfo{year}{2011}).

\bibitem[{\citenamefont{Ono et~al.}(1997)\citenamefont{Ono, Shimada, and
  Ootuka}}]{Ono:1997_a}
\bibinfo{author}{\bibfnamefont{K.}~\bibnamefont{Ono}},
  \bibinfo{author}{\bibfnamefont{H.}~\bibnamefont{Shimada}}, \bibnamefont{and}
  \bibinfo{author}{\bibfnamefont{Y.}~\bibnamefont{Ootuka}},
  \bibinfo{journal}{J. Phys. Soc. Jpn.} \textbf{\bibinfo{volume}{66}},
  \bibinfo{pages}{1261} (\bibinfo{year}{1997}).

\bibitem[{\citenamefont{Deshmukh and Ralph}(2002)}]{Deshmukh:2002_a}
\bibinfo{author}{\bibfnamefont{M.~M.} \bibnamefont{Deshmukh}} \bibnamefont{and}
  \bibinfo{author}{\bibfnamefont{D.~C.} \bibnamefont{Ralph}},
  \bibinfo{journal}{Phys. Rev. Lett.} \textbf{\bibinfo{volume}{89}},
  \bibinfo{pages}{266803} (\bibinfo{year}{2002}).

\bibitem[{\citenamefont{van~der Molen et~al.}(2006)\citenamefont{van~der Molen,
  Tombros, and van Wees}}]{vanderMolen:2006_a}
\bibinfo{author}{\bibfnamefont{S.~J.} \bibnamefont{van~der Molen}},
  \bibinfo{author}{\bibfnamefont{N.}~\bibnamefont{Tombros}}, \bibnamefont{and}
  \bibinfo{author}{\bibfnamefont{B.~J.} \bibnamefont{van Wees}},
  \bibinfo{journal}{Phys. Rev.} \textbf{\bibinfo{volume}{B 73}},
  \bibinfo{pages}{220406} (\bibinfo{year}{2006}).

\bibitem[{\citenamefont{Wunderlich et~al.}(2006)\citenamefont{Wunderlich,
  Jungwirth, Kaestner, Irvine, Wang, Stone, Rana, Giddings, Shick, Foxon
  et~al.}}]{Wunderlich:2006_a}
\bibinfo{author}{\bibfnamefont{J.}~\bibnamefont{Wunderlich}},
  \bibnamefont{et~al.}, \bibinfo{journal}{Phys. Rev. Lett.}
  \textbf{\bibinfo{volume}{97}}, \bibinfo{pages}{077201}
  (\bibinfo{year}{2006}).

\bibitem[{\citenamefont{Schlapps et~al.}(2009)\citenamefont{Schlapps, Lermer,
  Geissler, Neumaier, Sadowski, Schuh, Wegscheider, and
  Weiss}}]{Schlapps:2009_a}
\bibinfo{author}{\bibfnamefont{M.}~\bibnamefont{Schlapps}},
  \bibnamefont{et~al.},
  \bibinfo{journal}{Phys. Rev.} \textbf{\bibinfo{volume}{B 80}},
  \bibinfo{pages}{125330} (\bibinfo{year}{2009}).

\bibitem[{\citenamefont{Bernand-Mantel
  et~al.}(2009)\citenamefont{Bernand-Mantel, Seneor, Bouzehouane, Fusil,
  Deranlot, Petroff, and Fert}}]{Bernand-Mantel:2009_a}
\bibinfo{author}{\bibfnamefont{A.}~\bibnamefont{Bernand-Mantel}},
  \bibnamefont{et~al.},
  \bibinfo{journal}{Nat. Phys.} \textbf{\bibinfo{volume}{5}},
  \bibinfo{pages}{920} (\bibinfo{year}{2009}).

\bibitem[{\citenamefont{Thomson}(1857)}]{Thomson:1857_a}
\bibinfo{author}{\bibfnamefont{W.}~\bibnamefont{Thomson}},
  \bibinfo{journal}{Proc. Roy. Soc. London} \textbf{\bibinfo{volume}{8}},
  \bibinfo{pages}{546} (\bibinfo{year}{1857}).

\bibitem[{\citenamefont{McGuire and Potter}(1975)}]{McGuire:1975_a}
\bibinfo{author}{\bibfnamefont{T.}~\bibnamefont{McGuire}} \bibnamefont{and}
  \bibinfo{author}{\bibfnamefont{R.}~\bibnamefont{Potter}},
  \bibinfo{journal}{IEEE Trans. Magn.} \textbf{\bibinfo{volume}{11}},
  \bibinfo{pages}{1018} (\bibinfo{year}{1975}).

\bibitem[{\citenamefont{Gould et~al.}(2004)\citenamefont{Gould, {R\"{u}ster},
  Jungwirth, Girgis, Schott, Giraud, Brunner, Schmidt, and
  Molenkamp}}]{Gould:2004_a}
\bibinfo{author}{\bibfnamefont{C.}~\bibnamefont{Gould}},
  \bibnamefont{et~al.},
  \bibinfo{journal}{Phys. Rev. Lett.} \textbf{\bibinfo{volume}{93}},
  \bibinfo{pages}{117203} (\bibinfo{year}{2004}).

\bibitem[{\citenamefont{Tran et~al.}(2009)\citenamefont{Tran, Peiro, Jaffrès,
  George, Mauguin, Largeau, and Lemaître}}]{Tran:2009_a}
\bibinfo{author}{\bibfnamefont{M.}~\bibnamefont{Tran}},
    \bibnamefont{et~al.},
  \bibinfo{journal}{Appl. Phys. Lett.} \textbf{\bibinfo{volume}{95}},
  \bibinfo{pages}{172101} (\bibinfo{year}{2009}).

\bibitem[{\citenamefont{Jungwirth et~al.}(2006)\citenamefont{Jungwirth, Sinova,
  {Ma\v{s}ek}, {Ku\v{c}era}, and MacDonald}}]{Jungwirth:2006_a}
\bibinfo{author}{\bibfnamefont{T.}~\bibnamefont{Jungwirth}},
  \bibnamefont{et~al.}, \bibinfo{journal}{Rev. Mod. Phys.}
  \textbf{\bibinfo{volume}{78}}, \bibinfo{pages}{809} (\bibinfo{year}{2006}).

\bibitem[{Die(2008)}]{Dietl:2008_b}
vol.~\bibinfo{volume}{82} of \emph{\bibinfo{series}{Semiconductors and
  Semimetals}} (\bibinfo{publisher}{Elsevier}, \bibinfo{year}{2008}).

\bibitem[{\citenamefont{Pappert et~al.}(2006)\citenamefont{Pappert, Schmidt,
  {H\"{u}mpfner}, {R\"{u}ster}, Schott, Brunner, Gould, Schmidt, and
  Molenkamp}}]{Pappert:2006_a}
\bibinfo{author}{\bibfnamefont{K.}~\bibnamefont{Pappert}},
  \bibnamefont{et~al.},
  \bibinfo{journal}{Phys. Rev. Lett.} \textbf{\bibinfo{volume}{97}},
  \bibinfo{pages}{186402} (\bibinfo{year}{2006}).

\bibitem[{\citenamefont{Winkler et~al.}(2000)\citenamefont{Winkler, Papadakis,
  Poortere, and Shayegan}}]{Winkler:2000_a}
\bibinfo{author}{\bibfnamefont{R.}~\bibnamefont{Winkler}},
  \bibinfo{author}{\bibfnamefont{S.~J.} \bibnamefont{Papadakis}},
  \bibinfo{author}{\bibfnamefont{E.~P.~D.} \bibnamefont{Poortere}},
  \bibnamefont{and} \bibinfo{author}{\bibfnamefont{M.}~\bibnamefont{Shayegan}},
  \bibinfo{journal}{Phys. Rev. Lett.} \textbf{\bibinfo{volume}{85}},
  \bibinfo{pages}{4574} (\bibinfo{year}{2000}).

\bibitem[{\citenamefont{Dietl et~al.}(2001)\citenamefont{Dietl, Ohno, and
  Matsukura}}]{Dietl:2001_b}
\bibinfo{author}{\bibfnamefont{T.}~\bibnamefont{Dietl}},
  \bibinfo{author}{\bibfnamefont{H.}~\bibnamefont{Ohno}}, \bibnamefont{and}
  \bibinfo{author}{\bibfnamefont{F.}~\bibnamefont{Matsukura}},
  \bibinfo{journal}{Phys. Rev.} \textbf{\bibinfo{volume}{B 63}},
  \bibinfo{pages}{195205} (\bibinfo{year}{2001}).

\bibitem[{\citenamefont{Abolfath et~al.}(2001)\citenamefont{Abolfath,
  Jungwirth, Brum, and MacDonald}}]{Abolfath:2001_a}
\bibinfo{author}{\bibfnamefont{M.}~\bibnamefont{Abolfath}},
  \bibinfo{author}{\bibfnamefont{T.}~\bibnamefont{Jungwirth}},
  \bibinfo{author}{\bibfnamefont{J.}~\bibnamefont{Brum}}, \bibnamefont{and}
  \bibinfo{author}{\bibfnamefont{A.~H.} \bibnamefont{MacDonald}},
  \bibinfo{journal}{Phys. Rev.} \textbf{\bibinfo{volume}{B 63}},
  \bibinfo{pages}{054418} (\bibinfo{year}{2001}).

\bibitem[{\citenamefont{Zemen et~al.}(2009)\citenamefont{Zemen, Kucera,
  Olejnik, and Jungwirth}}]{Zemen:2009_a}
\bibinfo{author}{\bibfnamefont{J.}~\bibnamefont{Zemen}},
  \bibinfo{author}{\bibfnamefont{J.}~\bibnamefont{Kucera}},
  \bibinfo{author}{\bibfnamefont{K.}~\bibnamefont{Olejnik}}, \bibnamefont{and}
  \bibinfo{author}{\bibfnamefont{T.}~\bibnamefont{Jungwirth}},
  \bibinfo{journal}{Phys. Rev.} \textbf{\bibinfo{volume}{B 80}},
  \bibinfo{pages}{155203} (\bibinfo{year}{2009}).

\bibitem[{\citenamefont{Sawicki et~al.}(2005)\citenamefont{Sawicki, Wang,
  Edmonds, Campion, Staddon, Farley, Foxon, Papis, Kaminska, Piotrowska
  et~al.}}]{Sawicki:2004_a}
\bibinfo{author}{\bibfnamefont{M.}~\bibnamefont{Sawicki}},
  \bibnamefont{et~al.}, \bibinfo{journal}{Phys. Rev.}
  \textbf{\bibinfo{volume}{B 71}}, \bibinfo{pages}{121302}
  (\bibinfo{year}{2005}).

\bibitem[{\citenamefont{Shick et~al.}(2010)\citenamefont{Shick, Khmelevskyi,
  Mryasov, Wunderlich, and Jungwirth}}]{Shick:2010_a}
\bibinfo{author}{\bibfnamefont{A.~B.} \bibnamefont{Shick}},
  \bibnamefont{et~al.},
  \bibinfo{journal}{Phys. Rev. B} \textbf{\bibinfo{volume}{81}},
  \bibinfo{pages}{212409} (\bibinfo{year}{2010}).

\end{thebibliography}
\end{document}